# PATTERNS OF SOCIAL INFLUENCE
# IN A NETWORK OF SITUATED COGNITIVE AGENTS

Russell C. Thomas and John S. Gero

Krasnow Institute for Advanced Study
George Mason University
4400 University Drive, Mail Stop 2A1
Fairfax, VA, 22030, USA
e-mail: {rthoma12, jgero} @gmu.edu

## ABSTRACT

This paper presents the results of computational experiments on the effects of social influence on individual and systemic behavior of situated cognitive agents in a product-consumer environment. Paired experiments were performed with identical initial conditions to compare social agents with non-social agents. Experiment results show that social agents are more productive in consuming available products, both in terms of aggregate unit consumption and aggregate utility. But this comes at a cost of individual average utility per unit consumed. In effect, social interaction achieved higher productivity by 'lowering the standards' of individual consumers. While still at an early stage of development, such an agent-based model laboratory is shown to be an effective research tool to investigate rich collective behavior in the context of demanding cognitive tasks.

## INTRODUCTION

This paper investigates phenomenological patterns of collective behavior due to social influence among agents where agent value systems change in response to their experience with artifacts and their social interactions with other agents. This is the first phase of a larger research project to investigate innovation processes and policies across complex ecosystems of researchers, innovators, funding organizations, and consumers.

In general, systems of innovation can exhibit stability without convergence. They are capable of stable averages in aggregate activity but without stasis at a micro-level. By definition, systems of innovation produce a stream of new product types. Agents in the system must react and adapt to this stream of novelty, and in some sense master it.

The attention is on heuristic search processes when there is no clear global maximum in the landscape and there is no agent-accessible metric for collective utility, and where collective behavior is emergent. To emphasize the role of social interactions, agents are not endowed with spatial reasoning or spatial memory, nor are signaling or overt coordination capabilities provided. Search and learning is also hampered by local competition between consumers and also by a bias toward consuming to avoid frustration. These influences tend to cause agents to move away from local maxima even if they have found them. All together, it can be characterized as a 'frustrating search problem' for agents in that they might never be able to master the problem from an observer viewpoint.

In this paper, the theoretical lens on collective intelligence is situated cognition (Clancey 1997). Any cognitive system operates within its own worldview and that worldview affects its understanding of its interactions with its environment (Clancey 1997; Gero 2008). In essence, what you think the world is about affects what it is about for you.

A person or group of people is 'situated' because they have a worldview that is based on their experience. Situated cognition involves three ideas: situations, constructive memory and interaction. Situations are mental constructs that structure and hence give meaning to what is observed and perceived based on a worldview. Constructive memory makes memory a function of the situation and the past. Memory is not laid down and fixed at the time of the original sensate experience. What is remembered is constantly being recreated and reframed. Interactions between agents trigger changes in situations and memories.

Through the lens of situated cognition, collective intelligence is an emergent phenomenon that arises from the interplay of situations, constructive memory, and social interactions at the level of agents and networks of agents. Moreover, we believe that situated cognition is at the heart of social processes of creativity and inventiveness. This is why it is so important in the study of innovation (Gero 2011).

To facilitate this line of research, a computational laboratory using agent-based modeling with rich agents and rich artifacts is being built. The phenomena of interest arise through the dynamic social interactions between agents, and between agents and artifacts, far from equilibrium. Therefore, it is important that the system includes endogenous processes for learning (direct and indirect), social interactions and network formation, and be capable of rich emergent phenomena.

## ARCHITECTURE AND IMPLEMENTATION

### Functional Overview

There is one type of active agent in the current implementation of the system –Consumers – and one type of artifact – Products. Consumers are seeking to consume Products on a geographic Consumption Space with micro-behavior similar to foraging, but with social interactions. The Consumption Space is a bounded rectangular grid with von Neumann neighborhoods. In each clock cycle consumers can move to any neighboring point on the grid within the boundaries. Only one Consumer can occupy a given grid location at a given time. Figure 1 shows a visualization of Consumers and Products in the Consumption Space.

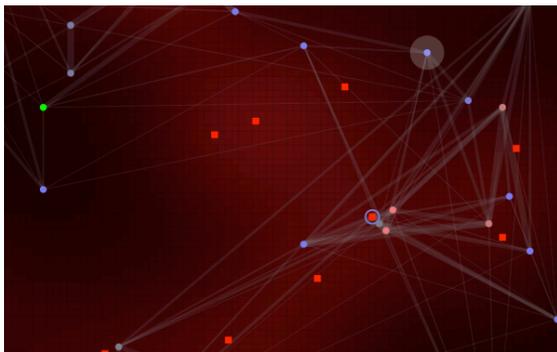

*Figure 1: Consumers and Products in Consumption Space (magnified).*

Decoding Figure 1: Products are red squares. Around each Product is a proximity gradient, which is a red-shaded area. This is an artifice to aide Consumer navigation. Consumers are shown as disks, color-coded according to their current consumption expectation: pink = optimistic; blue = pessimistic; grey = neutral; green = navigating toward another Consumer and not engaged in foraging. Consumers engaged in consumption are shown as circles around the Product they are consuming. Social ties are shown as an overlay network of translucent white lines, where the thickness of the line is proportional to the strength of the tie. In the upper right, the translucent disk over a Consumer indicates that the internal state of this Consumer is being monitored for diagnostic purposes.

Products have both a surface characteristics and functional characteristics. During their search and evaluation process, Consumers can only sense and perceive a Product's surface characteristics (its "signature"). The functional characteristics are only experienced through the process of consumption. During consumption, Consumers gain utility based on the functional characteristics relative to expectations. Higher than expected evaluations of functional characteristics yield positive utility, while lower than expected evaluations of functional characteristics yield negative utility. The surface characteristics of Products are related to their functional characteristics, but not identical. Consumers cannot directly perceive the value of products, though they can form expectations from past interactions.

The space of possible Product signatures, along with the utility of each Product, is called the Value Space. The value system for each Consumer centers on a single vector that represents the signature of its ideal product type. Consumers learn and adapt by adjusting this vector through experience and social interaction. Therefore, each Consumer's value system can be represented as a point in Value Space, and their changing values as paths through Value Space. All consumers have the same utility function that doesn't change during a run.

Consumers choose to consume based on their perception of a Product's signature, perception of its proximity to their ideal type, and a rough expectation of utility. Generally, Consumers choose to consume when the Product they encounter is close to their ideal type.

In summary, at the task level the Consumer's problem is to find relatively more desirable Products to consume by searching the Consumption Space and adjusting their ideal product type. If they become dissatisfied during this process or if they are not able to find products to consume, they interact socially to either modify their value system or to move toward another agent in the Consumption Space.

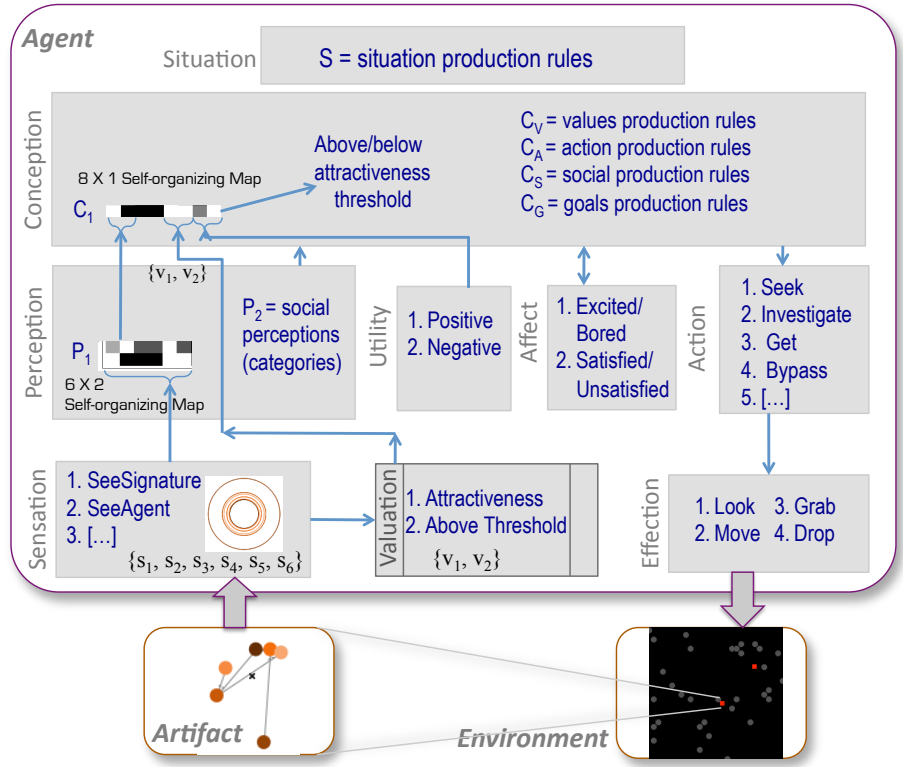

*Figure 2 Agent architecture*

At a social level, Consumers create and maintain social relationships through physical contact in the Consumption Space or through social interactions at a distance. However, if a Consumer is close to losing social connections, that Consumer interacts socially to build new connections through a referral process ('friends of friends').

**Architecture Overview**

*Consumers*

Figure 2 shows a block diagram of the agent architecture. Compared to other agent architectures in the social network influence literature (Friedkin and Johnsen 2011), this agent architecture is rich, in that includes both symbolic and sub-symbolic reasoning. This was necessary to implement situated cognition, which was one the primary research goals. For brevity only perception, conception, situation, and social interaction functions are described in the following paragraphs.

Perception is the collection of functions that enable the agent to focus and organize their sensations according to their current situation, their expectations, and past experiences. Consumers perceive Product signatures using a Self-organizing Map (SOM, also known as Kohonen Network).

SOMs are a type of neural network that are trained via unsupervised learning. Essentially they perform a mapping from the sensed Product Signature to a condensed 2D internal representation of the Products. This is functionally equivalent to conceptual spaces, as described in Gardenfors (2000). Perception is updated every clock cycle, but is only processed when new sensations arrive.

Perception of other agents is performed through a categorization and comparison procedure, where agents with direct social connections are labeled 'most similar', 'most dissimilar', 'most admired', 'least admired', etc.

Conception is the highest level of reasoning, including both tacit and explicit capabilities. Tacit conceptual reasoning is focused on deciding whether a given product should be considered attractive or not. This is implemented using a SOM that essentially creates a one-dimensional map of products that it has experienced and the value and/or utility that is perceived or realized from those products. A threshold value is used to trigger a decision that the product being inspected is attractive. The threshold value is adjusted through experience during the course of a run.

Conception also includes explicit reasoning about actions, social interactions and goals. These are implemented symbolically as production rules.

Situation has the function of cognitive orientation and focus. We implemented it using production rules that test for conditions that would change a Consumer's situation, and then fire actions to change their concepts according. Overlapping situations can be active at the same time. Situations act as conditions on other conception rules so that each conception rule fires only when one of its applicable situations is active. Situation also activates other reasoning functions, as appropriate.

We implemented the following generic situations:

| 1. `Consume-locally` | Local foraging |
|---|---|
| 2. `Interact-socially` | Interact with social neighbors. |
| 3. `Bored` | Decide what to change to end boredom |
| 4. `Dissatisfied` | Stop current actions, reverse expectations, and decide what to do instead |
| 5. `Change-location` | Use available information to pick a target destination |
| 6. `Change-values` | Make incremental changes in value system and selection criteria |
| 7. `Search-for-a-friend` | No social network, so find a friend |

Social interactions are implemented using production rules. Generally, social interaction only occurs when the Consumer is both not in the act of consumption and is also frustrated by its consumption experiences. The exception to this occurs when a Consumer's social ties have been reduced to two or fewer and their strength has fallen below a threshold value. Here, conception rules fire that cause the agent to create new social ties. This is necessary experimentally in order to sustain social networks and avoid disconnections. This was essential to maintain the distinction between "social" and "non-social" runs in the paired experiment design.

The targets of social actions and influences interactions are always defined by the perceptual categories mentioned earlier. In these experiments, Consumers are only influenced by one neighbor at a time. They do not poll their local social network or perform any reasoning based on the range of values of other agents.

The utility function for all agents is the same and is fixed. It is a logistic function of the number of edges in the Product structure (see below) relative to an expected value of 8. Edge counts above 8 yield positive utility while counts below yield negative utility.

The valuation function is also fixed and identical. It is the Euclidean distance between the Consumer's ideal product and the Product signature (6-element real valued vectors).

*Products*

Products are constructed as a graph with six vertices. During the construction process, edges between the vertices are formed at random, creating a single connected graph with between 5 and 15 edges. The Product's utility is a function of its topology, while its signature (external appearance) is a metric of its physical layout. Physical layout is constructed on a circle with a relaxation method to equalize the length of edges. Distance from the centroid of this layout produces a signature in the form of a 6-element vector.

This construction process produces a non-obvious relationship between the surface characteristics and the functional characteristics of Products. Products that are very close in surface characteristics (i.e. close in Value Space) may have very different utilities. This allows for rugged search landscapes, though any particular realization of 10 product types may or may not be rugged. When we generated a sample of 2000 product types, the resulting landscape was extremely rugged with no global maximum.

**Design Choices**

The following is an explanation of the most salient design choices we have made and their effect on system behavior and experiment results.

In contrast to other social influence networks research (e.g. Friedkin & Johnsen 1999), the Consumers are situated in two environments: the task environment of consumption and the social environment of agent-to-agent interaction. Thus, the social network is endogenous and dynamic and has a contextual influence over agent behavior in the task environment and their behavior on tasks influences their social world. This is a cornerstone in theory of situated cognition and, therefore, was a necessary choice given the objectives.

The agents are cognitive and adaptive rather than rational or even bounded rational. Specifically, the agents were not endowed with explicit preference ordering or utility maximization processes or capabilities. We believe the cognitive-adaptive model is more appropriate for innovation studies.

Deterministic rules for reasoning and acting are used whenever possible, especially for activities related to Product evaluation and consumption, and also related to social influence processes. This is in contradistinction with research approaches that use probabilistic rules for action decisions and influences. This has experimental benefits. The system is already endowed with several sources of randomness and mixing through agent interactions. Adding randomness into the agent architecture would have made it difficult to determine experimentally the cause-effect relationships. The system would have become essentially a large stochastic processes dominated by the Central Limit Theorem, producing statistically homogenous output.

Because the focus is on social influence involving value systems, Consumers have rather simple and reflexive capabilities for navigating in the Consumption Space. In contrast to insect or animal foraging models, the agents do not leave pheromone trails and don't use energy in the consumption process. We endowed the Consumption space with a proximity gradient for Products, so that moving toward or away from Products could be reduced to gradient ascent/decent. These choices are parsimonious because Consumer search in the space still resembles foraging, which was the intent. The main effect of this choice is that we need to control for density in the Consumption space so that it is neither too sparse nor too crowded.

## **EXPERIMENT DESIGN AND METHODS**

To evaluate the effect of social capabilities we ran a series of randomized paired simulations – social agents vs. non-social agents – and then used statistical methods to test hypotheses related to the differences in patterns of behavior. This experimental method is well suited to the objective of identifying significant and theoretically important differences for the system under test, given the particular nature of the agent architecture and the design choices described above. Readers should be cautious about assuming that the results apply generally to any social network of situated cognitive agents. Context matters.

### **Initial conditions**

The Consumption Space grid measured 165 X 165, and it was populated with 40 Consumers and 50 Products of 10 types with random uniform spatial distribution. All Consumers started with identical initial conditions except for their location in Consumption Space. Their ideal product signature was set to be the average of all product types. To initialize the SOMs for perception and conception, Consumers were primed by exposing them to each of the product types, allowing all of the reasoning functions to be performed on each product, followed by a learning step in between each evaluation. The social network between Consumers was initialized as a fully connected small world network, randomly generated using the Watts-Strogatz method.

For Products, we generate 10 types with a random construction technique, and retained types that were at least a minimum distance in their signatures. This minimum distance allowed us to maintain sufficient product variety and distribution in Value Space. In contrast, when we generated products using random construction rules with no constraints, the resulting distribution in Value Space tended to be clustered in one to three regions. While this isn't a difficultly in general, it did make agent cluster analysis difficult. By requiring minimum distances in Value Space for each product, we could be sure that the distribution of product types would allow Consumer clustering but not dictate it.

Each Product type was replicated five times to yield 50 Products. Even though Products are consumed during a run, the population of Products and types is static. This is accomplished by replacing the consumed Product in a new location at a random distance from its previous location, with Gaussian distribution.

Prior to running these experiments, we ran several tests over a range of parameters (number of Consumers, number of Products, size of Consumption Space, etc.). We found that the results described below were not sensitive to the number of Consumers or Products above 15 of each. However, the results were sensitive to the density of Consumers and Products in Consumption Space. Therefore, we chose parameter ratios that resulted in a targeted spatial density.

### **Experiment design**

The experiment consisted of 30 paired runs (social vs. non-social) of 10,000 clock cycles each. Pairs in a run were given the same random number generator seed, which created identical initial conditions for the pair, and also statistical independence between runs. Agent behavior and consumption results were tabulated every 20 clock cycles. For each period, we collected data on consumption and changes in Consumer value. We did not analyze Consumer paths through Consumption Space or the dynamics of the social network.

Test statistics were generally mean value for population aggregates or mean value for individuals in the population for a given metric.

## Statistical methods

To evaluate transients and trends in the time series, we performed linear regression and examined the slope parameter. Histograms and Q-Q plots were used to evaluate sample distributions, particularly to identify modes and deviations from Normality.

Hypothesis testing was done using the Paired T Test when sample distributions appeared Normal and Signed Rank Test when sample distributions did not appear Normal. In most cases we performed both tests.

To quantify the difficulty of the heuristic search problem, we used the Fitness Difficulty Correlation (FDC) from Jones & Forrest (1995). This metric is defined as follows.

Given a set $U = \{u_1, u_2, ..., u_n\}$ of $n$ individual utilities and a corresponding set $D = \{d_1, d_2, ..., d_n\}$ distances to the nearest local or global maximum, the correlation coefficient is

$$r = \frac{c_{UD}}{s_U s_D}$$

where

$$c_{UD} = \frac{1}{n}\sum_{i=1}^{n}(f_i - \overline{f})(d_i - \overline{d})$$

is the covariance of U and D, and $s_U, s_D, \overline{f}, \overline{d}$ are the standard deviations and means of U and D, respectively. The ideal landscape to search is one where utility is perfectly correlated to the inverse of distance, which is -1. This would be true for locally smooth landscapes. A very rough landscape would have FDC of zero.

## RESULTS

### Level of Difficulty

The following table summarizes the statistics for the difficulty of the heuristic search problem in the 30 experimental runs, using the Fitness Distance Correlation metric described above.

| FDC mean = -0.23 |
| --- |
| FDC standard deviation = 0.08 |

Because FDC is closer to 0 than -1, the results show that the level of difficulty was relatively high and that it was similar in all runs. The distribution of difficulty values appears to be Normal.

### Aggregate Consumption

We performed time series analysis of aggregate unit consumption for over individual runs for both social and non-social agents. The steady state trend is essentially flat, which we confirmed with linear regression, and the distribution of values is approximately Normal. In some runs there was a noticeable transient period of roughly 1,000 to 1,500 clock cycles. Since the experimental runs were 10,000, this transient did not affect the results described below.

Aggregate unit consumption was significantly higher for social agents than non-social agents. Figure 3 shows the smoothed kernel density estimate of the distribution of over 30 runs. Both the Paired T Test and Signed Rank Test reject the null hypothesis of no difference at the 0.1% significance level.

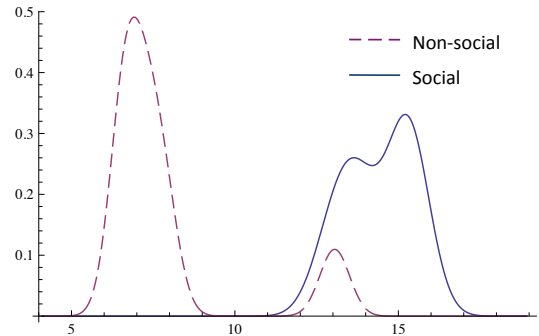

*Figure 3 Aggregate unit consumption means (kernel density estimate of pdf)*

The same relationship held for aggregate utility, also statistically significant at 0.1% level using the same tests. However, the higher level of aggregate utility for social agents was due to the higher rate of unit consumption. Average utility per unit consumed was lower for social agents compared to non-social agents.

### Behavior in Value Space

Another class of time series data is the movement of Consumer ideal product signatures (i.e. their "values") in Value Space. While all Consumers' values start at the same location in Value Space, we did not observe any cases of single point convergence as a steady state, either to global or local maxima in the Value Space, or to any other point. We did observe temporary clustering at or near the location of Products in Value Space, but neither social nor non-social Consumers showed any tendency toward long-term stability in clustering.

There was a statistically significant difference in the Value Space area covered by individuals during a run. Figure 4 shows the paired differences between social and non-social Consumers, where the difference is between mean areas covered by individuals in the population. Paired T Test and Signed Rank Test both reject the null hypothesis that there is no difference in means at the 1% significance level.

However, the histogram of paired differences is tri-modal. Investigation of individual runs indicates that this distribution structure is informative and not just due to random variation. For an analysis of this and possible implications, see the "Discussion" section, below.

We also examined path length in Value Space for individual agents. Figure 5 shows the smoothed kernel density estimate of means. Social Consumers had longer paths in Value Space than non-social Consumers. Paired T Test and Signed Rank Test both reject the null hypothesis that there is no difference at 0.1% significance level.

*a) Paired differences of means (histogram)*

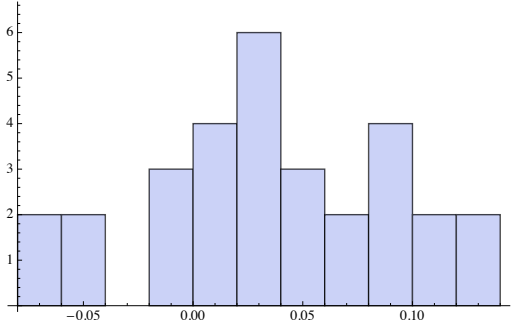

*b) Means*

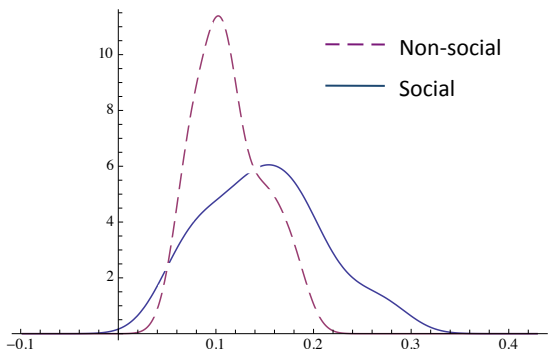

Figure 4 *Spatial coverage area in Value Space, population means (kernel density estimate of histogram)*

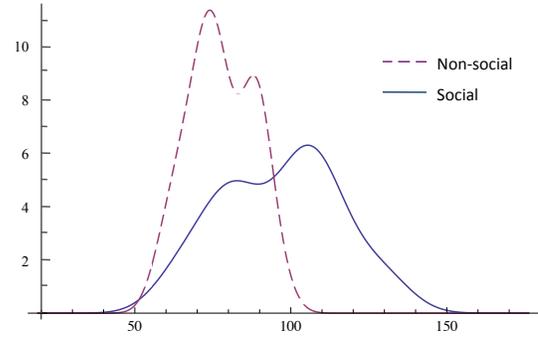

Figure 5 *Value Space coverage area means (kernel density estimate of histogram)*

Combining these two results, the data shows that social Consumers adjust their values more than non-social Consumers and that they adjust them over a wider range of values.

## DISCUSSION

The experiment results confirm that social interactions influence agent value systems in a way that changes their individual and collective performance in a frustrating task environment. Higher aggregate consumption for social agents is a result of moving individual values (ideal product signatures) away from favorable positions in Value Space. In essence, the social influence 'lowered the standards' of individuals that then allowed them to make more frequent consumption choices. Thus, aggregate consumption in units and utility was increased at the expense of average utility for individual Consumers.

The lack of clustering and lack of selectivity in Product consumption choices was surprising. But this goes to show that emergent phenomena cannot be always anticipated based on micro-level specifications. This is one of the virtues of computational models of social phenomena, and of rich agent-based models in particular. Neither social nor non-social Consumers were able to collectively optimize or converge on local or global maxima in Value Space. This result is a consequence of the design choices regarding the difficulty of heuristic search and also limitations on agent capabilities.

Trendless steady-state aggregate consumption rates were achieved in both cases, with approximately Normal distribution. Consistent with the design goals, Consumers did not reach any static equilibrium in terms of 1) values (i.e. their ideal product signature), 2) consumption pattern by Product type, or 3) social network structure. This indicates that in both cases the system was both stable and also far from equilibrium. This is important for the future research because we are interested in the emergence of equilibria or stable patterns that result from agent-agent interactions across agent types, and also agent-artifact interactions across different levels.

We believe the tri-modal histogram of Value Space coverage in Figure 4a merits further investigation. It appears that there are underlying relationships between the distribution of Products in Value Space and the behavior of social Consumers, specifically related to the formation of stable subgroups that specialize. This is an important result because it shows the potential for interesting and informative emergent phenomena in this type of computational system.

In summary, the experiment results support the assertion that social agents endowed with situated cognition have meaningful influence over the values of other agents and demonstrate forms of collective intelligence.

## FUTURE WORK

The future work will expand the functionality of the computational laboratory for the purpose of investigating innovation processes and policies in a complex multi-agent and multi-artifact ecosystem. The setting will be expanded to include dynamic population of Product types, both as exogenous shock of innovation and also endogenous production and invention processes by a new class of agents: Innovators. The functionality of Products will be enlarged so that they become capable of carrying and processing information, specifically to support endogenous modeling of research artifacts (i.e. intellectual property and funding proposals). The goal is explore innovation processes across a complex innovation ecosystem of agent and artifact types. When these features are fully developed, the aim is to experiment with innovation policies and to understand their collective behavior effects in systems of innovation.

## ACKNOWLEDGEMENTS

This research is supported by a grant from the National Science Foundation, grant no. SBE-0915482. Any opinions, findings, and conclusions or recommendations expressed in this material are those of the authors and do not necessarily reflect the views of the National Science Foundation.